\documentclass[manuscript]{aastex}  

\shorttitle{Charged particles time-dependent transverse transport}
\shortauthors{Fraschetti, Jokipii}

\usepackage{color}
\usepackage{amssymb,amsmath}

\begin{document}

\title{Time-dependent perpendicular transport of fast charged particles in a turbulent magnetic field}

\author{F. Fraschetti\altaffilmark{1,2} and J.R. Jokipii\altaffilmark{1}}
\affil{$^1$Departments of Planetary Sciences and Astronomy, University of Arizona, Tucson, AZ, 85721, USA}
\affil{$^{2}$LUTh, Observatoire de Paris, CNRS-UMR8102 and Universit\'e Paris VII,
5 Place Jules Janssen, F-92195 Meudon C\'edex, France.}

\begin{abstract}
We present an analytic derivation of the temporal dependence of the perpendicular transport coefficient of charged particles in magnetostatic turbulence, for times smaller than the time needed to  charged particles to travel the turbulence correlation length. 
This time window is left unexplored in most transport models. 
In our analysis all magnetic scales are taken to be much larger than the particle gyroradius, so that perpendicular transport is assumed to be dominated by the guiding center motion. Particle drift from the local magnetic field lines and magnetic field lines random walk are evaluated separately for slab and 3D isotropic turbulence. Contributions of wavelength scales shorter and longer than the turbulence coherence length are compared. In contrast to slab case, particles in 3D isotropic turbulence unexpectedly diffuse from local magnetic field lines; this result questions the common assumption that particle magnetization is independent on turbulence geometry. Extensions of this model will allow for a study of solar wind anisotropies.
\end{abstract}


\section{Introduction}

The behaviour of individual fast charged particles in magnetic turbulence is relevant to a number of problems in plasma astrophysics, from the solar wind \citep[e.g.][]{bc05} to interstellar medium \citep[e.g.][]{es04} and cosmic rays at highest energy \citep[e.g.][]{f08}. However, in contrast to cosmic rays with energies beyond the GeV scale, a thorough understanding of the particle transport properties can be attained only in interplanetary space, where {\it in situ} measurements of both magnetic turbulence energy spectrum and particles energy are possible. Diffusion theory \citep[e.g.][]{j66}, as a main tool to study charged particle propagation in magnetic turbulence, yields a statistical description of a population of particles and relies on the approximation \citep{j72} that a characteristic time $T$ exists much larger than the correlation time $t_c$ of the 
magnetic field fluctuations (as seen by the particle) but also much smaller than the time-scale of both the variation of these fluctuations and of the average distribution function. 
The Vlasov-Boltzmann equation for the charged particles phase-space distribution function can be therefore considerably simplified to terms of the second order moments of the magnetic field fluctuations. In this scenario higher-order moments are not necessary to determine the particles motion as the process is markovian; diffusion is governed by the central limit theorem \citep{c43}. 

Observational constraints posed by heliospheric environment on perpendicular diffusion across the average magnetic field involving, e.g. jovian electrons \citep{c77}, have not yet been included in a first-principles unified theoretical picture. Perpendicular diffusion occurring in the ecliptic plane is invoked as a plausible explanation of the time delays in solar energetic particle events detected by {\it Helios} \citep{wc06}. A more remarkable longitudinal separation in the combined electron observations by {\it Stereo A/B} and {\it SOHO} from the January 17, 2010 event \footnote{\footnotesize{Available at URL $http://www2.physik.uni-kiel.de/stereo/downloads/sept\_electron\_events.pdf$}} suggests a strong diffusion perpendicular to the mean magnetic field. The access of high energy particles to high-latitude heliospheric regions observed by {\it Ulysses} \citep{m09} is dominated by particle propagation along the mean magnetic field lines, although cross-field diffusion cannot be excluded. 

Perpendicular transport in a magnetic field depending on two or fewer space coordinates originates only from the meandering of the magnetic field lines \citep{jkg93} whereas in an arbitrary three-dimensional turbulence also emerges as a general property of the particle motion \citep{gj94}. Kinetic approach has been applied to perpendicular scattering of strongly magnetized charged particles by using a model for the collision integral \citep{cp93}. Recent numerical simulations \citep{mmbrb09} investigated the common assumption that the charged-particle gyrocenter follows the magnetic field lines: the approaching of the guiding center cross-field motion to the transverse field line random walk for various parallel mean free paths is studied for a solar-wind like turbulence. However we notice that the assumption that guiding center follows the magnetic field lines does not result directly from the equation of motion; therefore it may be realized only for particular turbulence models. 

An approximate diffusive perpendicular transport model based on the guiding center motion (Non-Linear Guiding Center, NLGC) has been put forward in \citet{mqbz03}, which provided a method to compute magnetic fluctuations along perturbed particle trajectory. However, the NLGC's assumption that the probability density of perpendicular displacement is diffusive at all times is a limitation of this model. Subdiffusive nature of perpendicular transport in a slab turbulence was not recovered, contrary to the expectation from the conservation of canonical momentum of ignorable coordinate \citep{jkg93} and to the findings from test particle numerical simulations in turbulence having a 
dominant slab component \citep{qmb02}. NLGC has been extended to early phase of perpendicular scattering \citep{lwsz10}, where the probability distribution function of the waiting-time between two scatterings decays slower than the exponential, therefore including memory effects (non markovian process). For a review of other improvements of NLGC see references in \citet{lwsz10}.

In this paper, we explore the transition to the diffusion regime in a magnetostatic turbulence by using the first-order orbit theory as proposed by \citet{ro70}, which is based on two assumptions: 1) the particle gyroradius is much smaller than any variation length scale of magnetic field and 2) the turbulent magnetic energy is much smaller than the average magnetic field energy.
This allows us to disentangle the cross-field particle motion into two separate components: field lines meandering and gradient/curvature drift from the local field line. In the present paper, the drift is meant to be the transverse 
gyroperiod-averaged motion of the guiding center away from the local field line \citep{ro70}.
Field-line meandering has been recognized in early solar turbulence studies to be the main contribution to perpendicular motion to average field direction \citep{j66,jp68}; the diffusive nature of the field line spread has been related to the turbulence power spectrum power law index at low wavenumber \citep{r99}. On the other hand, individual particle gradient/curvature drift on scales smaller than correlation scale does not seem to have been object of theoretical investigation (see however \citet{sj10}), on the basis of the common belief that particle magnetization does not depend on the particular turbulence geometry, in contrast to recent numerical findings \citep{ts10}
Therefore, the knowledge of particle trajectory along and across local field lines has remained completely undetermined. In this paper we shed light to this distinction analyzing cases propedeutically relevant to the solar wind propagating cosmic rays and astrophysical blast waves of supernova remnants.

\section{Transverse guiding center drift}

We consider a process of propagation of a charged particle in magnetic turbulence which is statistically homogeneous in time, i.e., the velocity correlation depends only on time difference along the orbit. The instantaneous mean square displacement along the space coordinate $x$ after a time $\Delta t = t - t'$ for a particle propagating in an arbitrary medium can be defined as \citep{t22,g51,k57} 
\begin{equation}
{ \langle (\Delta x) ^2\rangle } = 2 \int_0^{\Delta t} d\xi (\Delta t - \xi)   \langle v_x(t' + \xi) v_x(t')\rangle  \, ,
\label{X}
\end{equation}
where $t'$ is an arbitrary initial time, $\xi$ the time lag and the ensemble average $\langle ..\rangle$ is meant to be the average over a population of particles and over an ensemble of turbulence realizations.
We notice that Eq.(\ref{X}) applies for any value of $\Delta t$. We define
\begin{equation}
d_{xx} (t) \equiv \frac{1}{2} \frac{d}{d t} { \langle (\Delta x) ^2\rangle }  =  \int_0^{ t} d\xi  \langle v_x(t' + \xi) v_x(t')\rangle \, .
\label{dxx}
\end{equation}
The standard perpendicular coefficient of diffusion can then be defined as 
\begin{equation}
\kappa_{xx} = \displaystyle\lim_{ t \rightarrow \infty} d_{xx} ( t).
\end{equation}
We here investigate the transverse motion of a low-rigidity particle for a time smaller than $t_c$ in general 3D magnetostatic turbulence. The diffusion approximation therefore may not be valid. 

We consider a spatially homogeneous, fluctuating, time-independent magnetic field. The amplitude of the fluctuation (${\delta B}$) is assumed to be much smaller than the average field magnitude ($B_0$).
We represent such a magnetic field as ${\bf B(x) = B}_0 + \delta {\bf B(x)}$, with an average component ${\bf B}_0 = B_0 {\bf e}_z$ and $\langle \delta {\bf B(x)} \rangle = 0$ and $\delta B({\bf x})/B_0 \ll 1$. This approximation is known to be valid in several turbulent media, as the solar wind, where the propagation of the magnetic fluctuation is much smaller than the velocity of the bulk ionized fluid. We will make use of the first-order orbit theory \citep{ro70}: the particle gyroradius $r_g$ is much smaller than the length-scale of any magnetic field variation:
\begin{equation}
r_g \ll \displaystyle\min_{i,j=1,3} \left |\frac{B_i}{\partial_j B_i}\right|  \, , 
\label{first_order}
\end{equation}
where $B_i$ is the {\it i}-th component of the perturbed field ${\bf B (x)}$.
No further assumption is made on the spatial dependence of $\bf \delta B$ or geometry.
In this approximation, we consider the guiding center motion. In a spatially varying magnetic field, the guiding center may significantly drift from the average field direction due to the action of the field gradient on the particle magnetic moment. We therefore consider non-zero gradient and curvature drifts. We estimate that drift and resulting displacement after a time shorter than the correlation time. The guiding position ${\bf X} (t) = (X, Y , Z)$ for a particle of mass $m$, charge $Ze$ 
and momentum $\bf{p}$ having coordinate ${\bf x} (t) = (x, y, z)$ is described, in c.g.s. units, by
\begin{equation}
{\bf X} (t) = {\bf x} (t) - \frac{c}{Ze}\frac{{\bf B}\times{\bf p}(t)}{B^2} \, .
\label{GCmotion}
\end{equation}
If the scales of magnetic fluctuation are much larger than the gyroradius $r_g$, the guiding center motion defined in 
Eq.~(\ref{GCmotion}) has the role of effective gyroperiod-averaged motion. 
Therefore, Eq.~(\ref{GCmotion}) can well describe the motion perpendicular to the local magnetic field.
In the case of ``finite Larmor radius'', the gyroradius only represents the typical scale of particle motion and Eq.~(\ref{GCmotion}) provides the instantaneous guiding center position whereas 
gyroperiod-average becomes meaningless.
In the magnetostatic field described above, the guiding center velocity transverse to the field ${\bf B (x)}$ is given at the first order in $\delta B({\bf x})/B_0$ by the gyroperiod average \citep{ro70}
\begin{eqnarray}
{\bf V}_{\perp}^G (t) =  \frac{vpc}{Ze B^3} \left[ \frac{1+\mu^2}{2} {\bf B}\times \nabla B + \mu^2 B (\nabla \times {\bf B})_\perp   \right]
\label{Vperp}
\end{eqnarray}
where $\alpha$ is the particle pitch angle and $\mu = {\rm cos}\alpha$. Eq.~(\ref{Vperp}) gives the first order most general expression of the guiding center velocity orthogonal to the local magnetic field direction \citep{b88}. Here the variation of $\alpha$ is assumed to be negligible over a gyroperiod.  
Being ${\bf V}_{\perp}^G (t)$ a gyroperiod average, magnetic field can be computed at the guiding center position during that gyroperiod. In contrast to \citet{mqbz03}, the transverse motion of the guiding center from the field line is not parametrized in the present paper through some constants to be inferred from numerical simulations, but described directly from the equation of motion of the guiding center.
The finite-time average square transverse displacement of the particle from the direction of local $B$ due to drift $d_D (t)$ can then be written in this approximation using the Eq.(\ref{dxx}):
\begin{equation}
d_{D_{ii}} (t) = \int _0 ^t  d\xi \langle {\bf V}_{\perp, i}^G (t') {\bf V}_{\perp, i}^G (t' + \xi) \rangle
\label{dXX1}
\end{equation}
where $i$ stands for any transverse coordinate, $X$ or $Y$.
The average square displacement is computed from the following expression, to the lowest order in $\delta B/B_0$,
\begin{eqnarray}
\lefteqn{d_{D_{ii}}  (t) \simeq \left(\frac{vpc}{Ze B_0 ^2} \right)^2 \times }\nonumber \\ 
& & \int_0 ^t   d\xi \langle \left [ \frac{1 - \mu^2}{2}\partial_j \delta B_3 + \mu^2 \partial_3 \delta B_j \right][{\bf x}(t')] \times \nonumber \\
& & \left [ \frac{1 - \mu^2}{2}\partial_j \delta B_3 + \mu^2 \partial_3 \delta B_j \right][{\bf x}(t' + \xi)]  \rangle \; ,
\label{dXX}
\end{eqnarray}
where $(i,j) = (1,2)$ or $ (2,1)$ and the fields are evaluated at the perturbed particle position ${\bf x}(t)$. The average square of the displacement $d_{D_{ii}}  (t)$ does not depend on the sign of the electric charge $Ze$, at variance from the drift velocity in Eq.~(\ref{Vperp}). We notice that in the first-order orbit approximation the particular case of 2D turbulence defined by $\delta B (x,y) = (\delta B_x, \delta B_y, 0)$ provides a zero transverse velocity drift; this is because, on the right hand side of Eq. (\ref{dXX}), this form of turbulence has $\delta B_3 = 0$ and also $\delta B_j$ does not depend on the z coordinate. Thus, this analytic method cannot be applied to the composite slab/2D solar wind model of \citet{bwm96}, a very useful but empirical description of the MHD-scale turbulence in slow solar wind. Moreover, the slab/2D model could be incomplete as the non-wave (``2D'') turbulence might have an additional component along ${\bf B}_0$. This implies that, due to the sub-diffusive nature of the perpendicular particle transport in slab turbulence, drifts from local field-line found in numerical simulations for composite model \citep{mmbrb09} are second-order contributions. Different anisotropies may be compatible with large-scales solar wind observations; 
in this paper we indicate a possible alternative method. 

We may simplify the derivation by using the Fourier representation of $\delta {\bf B}({\bf x})$:
\begin{equation}
\delta {\bf B}({\bf x}) = {\Re} \int_{-\infty} ^{\infty} d^3 k \, {\bf \delta B(k)} e^{i {\bf k \cdot x} (t)} 
\label{B_Fourier}
\end{equation}
where ${\Re} (\cdot)$ stands for the real part and ${\bf  x} (t)$ is the particle position at time $t$.
Therefore the average displacement in Eq.(\ref{dXX}) contains terms of type
\begin{equation}
\partial_l \delta B_j ({\bf x}) = {\Re} \int_{-\infty} ^{\infty} d^3 k \, \delta B_j({\bf k}) (ik_l) e^{i {\bf k \cdot x} (t)} 
\label{lj_1}
\end{equation}
with $l,j=1,2,3$. We compute the particle position in Eq.(\ref{lj_1}) along the local magnetic field: ${\bf x} (t) = {\bf x}_0 (t) + {\bf x}_{MFL} (z(t))$, where the unperturbed particle orbit is ${\bf x}_0 (t) = (v {\rm sin}\phi \sqrt{1-\mu^2}/\Omega, - v {\rm cos}\phi \sqrt{1-\mu^2}/\Omega, v_\parallel t)$; here $v_{\parallel}$ is the unperturbed particle velocity along $z$, $\phi$ is the particle azimuth angle in the plane orthogonal to ${\bf B}_0$ and $\Omega = ZeB_0/(m\gamma c)$ the particle gyrofrequency in the background field containing the Lorentz factor $\gamma$; ${\bf x}_{MFL} (z(t)) = (x_{MFL}, y_{MFL}, z(t))$ is the offset in the plane orthogonal to $B_0$ due to the magnetic field line random walk (MFLRW) at $z=z(t)$. The assumption of ballistic motion along $B_0$, i.e., $z=v_\parallel t$, relies on the choice $\delta B \ll B_0$; at times smaller than the correlation time of the perpendicular fluctuation, {\it a fortiori} we cannot assume parallel diffusion. At the small length-scales considered here, parallel and perpendicular motions can be disentangled and any non-markovian parallel motion, e.g., memory effect of a particle tracing back its trajectory, is not expected to interfere with the perpendicular transport, in contrast to the case of compound diffusion.  We can write
\begin{equation}
e^{i {\bf k \cdot x} (t)}  \simeq  e^{i {\bf k \cdot x}_0 (t)} e^{i {\bf k \cdot x}_{MFL} (z(t))} \, .
\label{exp0}
\end{equation}
The magnetic field lines (MFL) are defined by $d {\bf x}_{MFL} \times {\bf B} = 0$. This implies that a finite distance $\Delta {\bf x}_{MFL}$ in the ballistic approximation is a first order term in $\delta{\bf  B}$: $\Delta {\bf x}_{MFL}  \simeq (\delta {\bf B}/B_0) v_\parallel t$. Therefore, the exponential $e^{i {\bf k \cdot x}_{MFL}  (z(t))}$ contribute only at zero order in Eq.~(\ref{exp0}) and the fuctuation in Eq.~(\ref{lj_1}) can be computed along the unperturbed trajectory: $e^{i {\bf k \cdot x} (t)}  \simeq  e^{i {\bf k \cdot x}_0 (t)}$, which is equivalent to the quasi-linear approximation.
To first order in Eq.(\ref{lj_1}) we can replace the exponential as
\begin{equation}
e^{i {\bf k \cdot x} (t)}  \simeq  e^{i (W{\rm sin}(\psi - \phi) + k_\parallel v_\parallel t)} \, ,
\label{exp0_2}
\end{equation}
where $\psi = {\rm tg}^{-1} (k_y/k_x)$, $k_\parallel = k_z$, $k_\perp = \sqrt{k_x^2 +k_y^2}$ and $W = k_\perp v \sqrt{1-\mu^2}/\Omega = k_\perp r_g$. 

By using the Bessel function identities (see \citet{as64}, Eq. (9.1.41))
\begin{equation}
e^{i z {\rm sin}\phi} = \sum_{n=-\infty} ^{\infty} J_n (z) e^{i n \phi} \, ,
\end{equation}
Eq.(\ref{exp0}) is rewritten as (see also \citet{s02}, Sect. 12.2.1):
\begin{equation}
e^{i {\bf k \cdot x_0}(t)} = \sum_{n=-\infty} ^{\infty} J_n (W) e^{ik_\parallel v_\parallel t + in(\psi - \phi + \Omega t)} \, .
\label{exp}
\end{equation}
The magnetic fluctuation space derivative in Eq.(\ref{lj_1}) can be then written in the following way
\begin{eqnarray}
\partial_l \delta B_j ({\bf x}) & = & {\Re}   \sum_{n=-\infty} ^{\infty} \int_{-\infty} ^{\infty} d^3 k \, {\delta B_j(k)}  (ik_l) J_n (W) \times \nonumber \\ 
 &  & e^{ik_\parallel v_\parallel t + in(\psi - \phi + \Omega t)} 
\label{lj}
\end{eqnarray}
with $l,j=1,2,3$. The typical term in Eq.(\ref{dXX}) is of type 
\begin{equation}
{\Re} \left(\frac{vpc}{Ze B_0 ^2} \right)^2 \,F(\mu^2) \int_0 ^t d\xi \langle  \partial_l  \delta B_r [{\bf x}(t')] \cdot \partial_p \delta B_q^*[{\bf x}(t' + \xi)]  \rangle \, ,
\label{product}
\end{equation}
here $F(\mu^2)$ represents various $\mu$ factors resulting from the expansion of Eq.(\ref{dXX}).
Using Eq.(\ref{lj}), we obtain for Eq.(\ref{product})
\begin{eqnarray}
\lefteqn{{\Re} \left(\frac{vpc}{Ze B_0 ^2} \right)^2 \,F(\mu^2) \times} \nonumber \\ 
& & \int_0 ^t d\xi \langle  \sum_{n=-\infty} ^{\infty} \sum_{m=-\infty} ^{\infty}  \int_{-\infty} ^{\infty} d^3 k  \int_{-\infty} ^{\infty} d^3 k' \delta B_r({\bf k}) \times \nonumber \\  
& & J_n (W) (ik_l)  \delta B^*_q({\bf k}') J^*_m (W) (-ik'_p) \times \nonumber\\ 
& & e^{[i(k_\parallel -k'_\parallel) v_\parallel t' + i(n-m)(\psi - \phi + \Omega t')  - i(k'_\parallel v_\parallel + m \Omega)\xi]} \rangle  \, .
\label{expl}
\end{eqnarray}
We assume the standard inertial range magnetic turbulence power spectrum which is uncorrelated at different wavenumber vectors:
\begin{equation}
\langle \delta B_r({\bf k}) \delta B_q ^*({\bf k'})  \rangle = \delta ({\bf k} - {\bf k'}) P_{rq}({\bf k}) \,  .
\label{PS}
\end{equation}
Thus Eq.(\ref{expl}) reduces to
\begin{equation}
{\Re} \,\left(\frac{vpc}{Ze B_0 ^2} \right)^2 F(\mu^2) \sum_{n=-\infty} ^{\infty} \int_{-\infty} ^{\infty} d^3 k R({\bf k}, t) P_{rq}({\bf k}) k_l k_p J_n ^2 (W)
\label{Fmu}
\end{equation}
whose time-dependence is entirely contained in
\begin{equation}
R({\bf k}, t) \equiv \int_0 ^t d\xi e^{-i(k_\parallel v_\parallel + n \Omega)\xi} =
 \frac{e^{-i(k_\parallel v_\parallel + n  \Omega)t} - 1}{-i(k_\parallel v_\parallel + n  \Omega) } \, .
\label{Rt}
\end{equation}
Since ${\Im }J_n (W) = 0$, where ${\Im }(\cdot)$ stands for imaginary part, we may consider ${\Re} R({\bf k}, t)$:
\begin{equation}
{\Re} R({\bf k}, t) = \frac{{\rm sin} [(k_\parallel v_\parallel + n  \Omega)t]}{k_\parallel v_\parallel + n  \Omega} \, .
\label{ReR}
\end{equation}
The orthogonal scale $1/k_\perp$ can be estimated as $|{B_i}/{\partial_j B_i}|$, thus Eq.(\ref{first_order}) states
\begin{equation}
W \sim k_\perp v/\Omega \ll1 \, .
\end{equation}
For $W \ll1$, it holds $J_0(W) \gg J_n(W)$ for $n \geq 1$; moreover, ${\Re} R({\bf k}, t) \sim 1/n$ for large $n$. We may therefore approximate the sum in Eq.(\ref{Fmu}) as its term with $n=0$. Therefore Eq.(\ref{Fmu}) yields, using Eq.(\ref{ReR}), four terms of type:
\begin{equation}
\left(\frac{vpc}{Ze B_0 ^2} \right)^2 F(\mu^2) \int_{-\infty} ^{\infty} d^3 k  P_{rq}({\bf k}) k_l k_p \frac{{\rm sin} [k_\parallel v_\parallel t]}{k_\parallel v_\parallel} \, ,
\label{Fmu2}
\end{equation}
with indexes $(r,q,l,p) = (3,3,2,2)$, $(3,2,2,3)$, $(2,3,3,2)$, $(2,2,3,3)$ for $d_{D_{XX}}$ and $(r,q,l,p) = (3,3,1,1)$, $(3,1,1,3)$, $(1,3,3,1)$, $(1,1,3,3)$ for $d_{D_{YY}}$. Eq. (\ref{Fmu2}) represents the general term contributing to the first-order transverse drift coefficient of a particle in a static first-order perturbed magnetic field.

\section{Magnetic-field-line random walk}

In the present section we compute the contribution to the time-dependent particle transverse transport due to MFLRW. If the correlation function of the magnetic fluctuation is homogeneous in space, the mean square displacement of the MFL orthogonal to the z-axis can be defined, in analogy to Eq.(\ref{dxx}), as 
\begin{eqnarray}
d_{MFL} (z) & \equiv & \frac{1}{2} \frac{d}{dz} { \langle (\Delta x_{MFL}) ^2\rangle } (z) \nonumber\\
& = & \frac{1}{B_0^2}\int_0^z d z'  \langle \delta B_x [{\bf x}(z')] \delta B_x [{\bf x}(0)] \rangle \, .
\label{dMFL}
\end{eqnarray}
We compute the magnetic turbulence $\delta {\bf B} ({\bf x})$ in Eq.(\ref{dMFL}) along the unperturbed trajectory of a particle travelling with zero pitch-angle, as in Eqs.(\ref{B_Fourier}, \ref{exp0_2}). The motion along the average field is then ballistic, i.e. $z=v_\parallel t$. In these approximations the transverse displacement of a MFL corresponding to a distance $v_\parallel t$ along $B_0$ travelled by a small rigidity particle can be written as 
\begin{equation}
d_{MFL} (t) =  \frac{1}{B_0^2} \int_{-\infty} ^{\infty} d^3 k  P_{rq}({\bf k}) \frac{{\rm sin} [k_\parallel v_\parallel t]}{k_\parallel} \, .
\label{dMFL2}
\end{equation}
Equation~(\ref{dMFL2}) is in agreement with Eq. (17) of \citet{s05} derived for pure slab turbulence.
The MFL coefficient diffusion describing the random walk of the field lines can then be defined as 
\begin{equation}
\kappa_{MFL} = \displaystyle\lim_{t \rightarrow \infty} d_{MFL} (t).
\end{equation}
In our approach, MFL diffusion cannot be assumed because travelled distance $z$ smaller than parallel correlation lengths is considered;
nevertheless, a simple ballistic motion along the z-axis allows to recover the standard result of MFL perpendicular diffusion in QLT. The discussion of the previous two section implies that the instantaneous coefficient diffusion perpendicular to the average magnetic field $B_0$ is given, in the presence of weak turbulence and neglecting parallel scattering, by two contributions: the random walk of the field line and the guiding center drift from the field line:
\begin{eqnarray}
d(t) & = &  d_D (t) + v_\parallel d_{MFL}(t),  \nonumber\\
  \kappa & = & \kappa_D +v_\parallel  \kappa_{MFL} = \displaystyle\lim_{t \rightarrow \infty} [d_D (t) + v_\parallel d_{MFL}(t)] .
\end{eqnarray}
In the next section the previous results are applied to the slab and 3D isotropic turbulences.

\vspace{-1cm}
\section{The turbulence power spectrum}

In this section we apply the approach developed in previous sections to derive the instantaneous transverse particle transport coefficients both of guiding center drifting from local MFL (see Eq.(\ref{Fmu2})) and of the MFL from the average field direction in Eq.(\ref{dMFL2}) by using the coherence length of the turbulence to disentangle small from large scale contributions to perpendicular diffusion. We will consider two cases: 1) slab turbulence, introduced \citep{j66} to represent the static limit of the solar wind magnetic fluctuations and extensively studied with Monte Carlo numerical simulations; we will compare the result with the QLT limit; 2) 3D isotropic turbulence, idealized case likely to provide an unperturbed model for anisotropies observed in the solar wind. 

\vspace{-1cm}
\subsection{Slab}\label{slab}

We consider the slab turbulence, static limit of transverse and longitudinal-propagating Alfven waves: $\delta {\bf B} = \delta {\bf B} (z)$ and ${\bf \delta B(x) \cdot e}_z = 0$. In this case, from Eq.(\ref{dXX}), $d^s _{D_{XX}}(t) = d^s _{D_{YY}}(t) = d^s _D (t)$. The turbulence wave number is aligned to the average magnetic field, thus we adopt the following form of the power spectrum:  $P_{rq}({\bf k}) = G(k_\parallel) (\delta(k_\perp)/k_\perp) \delta_{rq}$ with $r,q = 1,2$, and $P_{3i}({\bf k})=0$ with $i=1,2,3$. The 1D spectrum is assumed to be of Kolmogorov type. Observations of electron-density fluctuations inferred from scintillation measurements exhibit a Kolmogorov power law, with index approximately equal to 5/3, over 5 orders of magnitude \citep{a95}. Several other observations of magnetic turbulent media, from earth's magnetosphere to galaxy clusters, validate the Kolmogorov power spectrum up to a range of 12 orders of magnitude. We mention that solar wind observations show that at scales smaller than the ion thermal gyroradius ($\sim 10^7$ cm around the earth), much smaller than the scales considered in this paper, the magnetic turbulence spectrum deviates from the Kolmogorov, having an index of $-2.12$ \citep{b05}. At length-scales larger than the coherence length the measured interplanetary magnetic turbulence is well described by a flattening power spectrum \citep{h75,bms94}. On the other hand, a consistent comparison with the quasi-linear limit requires the power spectrum to be defined at scales larger than coherence length, i.e. for $k_\parallel < k_\parallel^{min}$, up to the physical scale of the system $2\pi/k_\parallel ^0$; we will adopt here a simplified form:
\begin{equation}
 G( k_\parallel)= \left\{
  \begin{array}{cc}
    G_\parallel ^0 k_\parallel ^{-q} & \mathrm{if~} k_\parallel ^{min} < k_\parallel  <  k_\parallel ^{max}  \, \\
    G_\parallel ^0 (k_\parallel^{min}) ^{-q} & \mathrm{if~} k_{\parallel} ^0< k_\parallel  <  k_\parallel ^{min}\, ,
    \label{spectrum}
   \end{array}
\right.
\end{equation}
where $k_\parallel ^{max}$ corresponds to the scale where the dissipation rate of the turbulence overcomes the energy cascade rate. The choice of a constant power spectrum at large scales instead of a function smoothly  connected to the inertial range already used in the literature is merely dictated by easier mathematical tractability.
Here $q=5/3$ and the constant $G_\parallel ^0$ is determined from the normalization
\begin{equation}
(\delta B)^2 = \int_{-\infty} ^{\infty} d^3 k (P_{11} + P_{22} +  P_{33}) 
\label{norma1}
\end{equation}
implying, using cylindrical coordinate ($d^3 k = dk_\parallel k_\perp dk_\perp d\psi$),
\begin{equation}
G_\parallel ^0 = \frac{(\delta B)^2 (q-1)}{ 4\pi q (k_\parallel^{min})^{1-q}} \, ,
\end{equation}
with the assumption $k_{\parallel} ^0 \ll k_\parallel ^{min} \ll k_\parallel ^{max}$. 

\begin{figure}
\includegraphics[width=10cm]{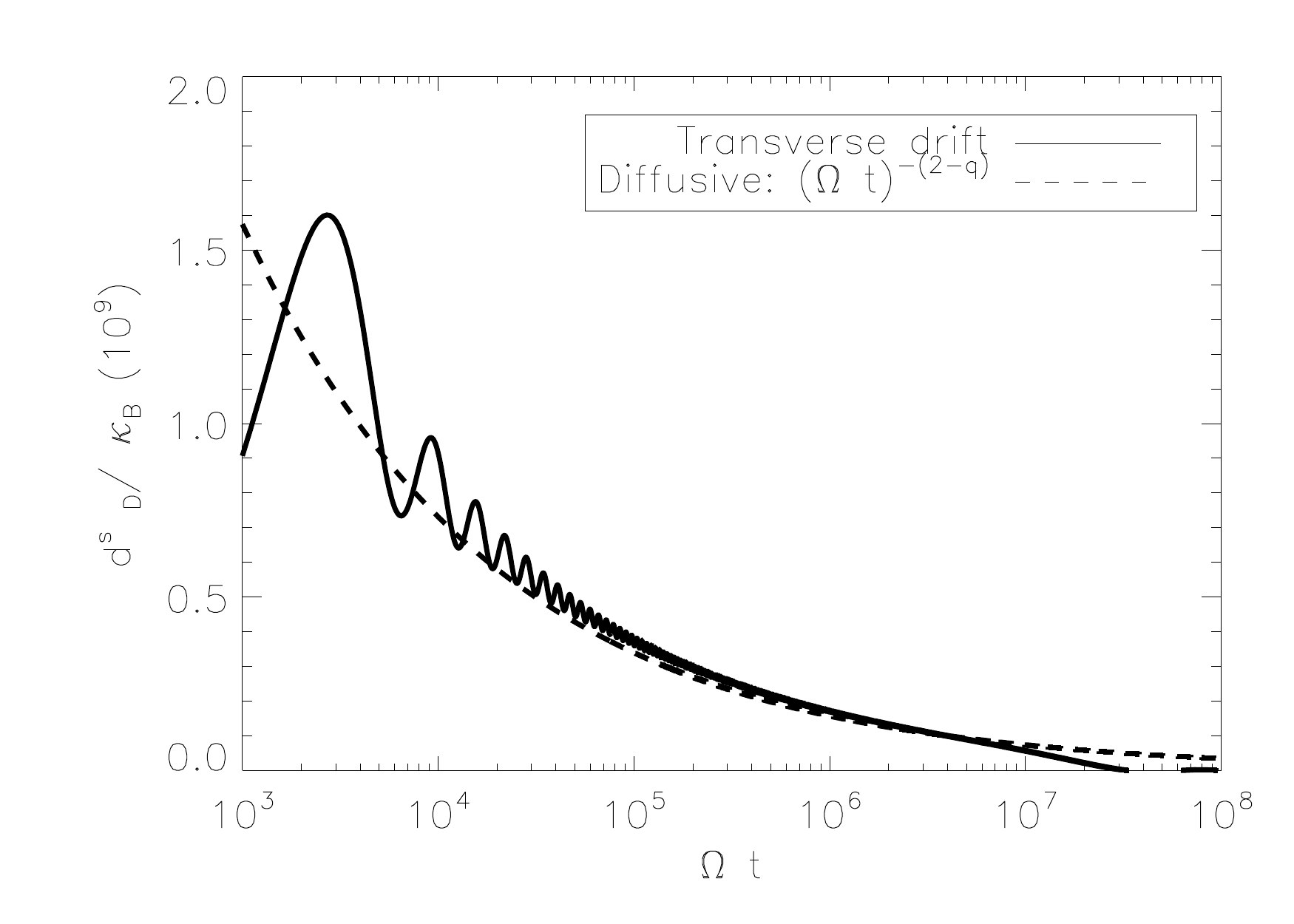}
\caption{Average transverse drift coefficient in units of $\kappa_B$ (rescaled by a factor $10^{9}$) as a function of $\Omega t$ for $k_{\parallel} ^{max} r_g= 10^{-3}$, $k_{\parallel} ^{max}/ k_{\parallel} ^{min} =10^4$, $k_{\parallel} ^{min}/k_{\parallel} ^0 =10^2$, $q=5/3$ and $\delta B/ B = 0.1$. The diffusive approximation in Eq. (\ref{dXXlimit2}) is superposed ({\it dashed line}). The departure from the perturbed field line of the guiding center, which in the first-order orbit theory represents the real particle motion with a good approximation, decreases to zero. Thus in a slab turbulence charged particles are tied to the weakly perturbed magnetic field lines. \label{drift}}
\end{figure}

We consider first the transverse drift in Eq.(\ref{Fmu2}). We average $F(\mu^2)$ over an isotropic pitch angle distribution. Using cylindrical coordinate ($d^3 k = dk_\parallel k_\perp dk_\perp d\psi$) we have
\begin{equation}
d^s _D (t) =  \left(\frac{vpc}{Ze B_0 ^2} \right)^2 \frac{\pi}{5} \int_{k_\parallel ^0}^{k_\parallel^{max}}d k_\parallel k_\parallel ^{2} G(k_\parallel) \frac{{\rm sin} [k_\parallel v_\parallel t]}{k_\parallel v_\parallel} 
\label{dXXlimit}
\end{equation}
In units of the Bohm coefficient diffusion ($\kappa_B = (1/3)r_g v$) and approximating $r_g/v_\parallel \simeq \Omega^{-1}$, we obtain

\begin{equation}
\frac{d^s _D (t)}{\kappa_B}  =   {3\over 20} \left( \frac{\delta B}{B_0} \right)^2 \frac{ q-1}{q} F(y_\parallel^m, y_\parallel^0, q)
\label{dXXlimit2}
\end{equation}
where we defined
\begin{equation}
F(y_\parallel^m, y_\parallel^0, q) = k_\parallel^{min} r_g \left[\frac{I(2-q, y_\parallel ^m)}{(y_\parallel^m)^{2-q}} + \frac{\sin y_\parallel - y_\parallel \cos y_\parallel}{y_\parallel ^2}  \Bigl\lvert^{y_\parallel^{m}} _{y_\parallel^0}  \right]
\label{F1}
\end{equation}
where the time-dependence is contained in the new variable $y_\parallel ^{m} = k_\parallel ^{min} v_\parallel t \simeq k_\parallel ^{min}  r_g \Omega t$ (and $y_\parallel ^{0} = k_\parallel ^{0} v_\parallel t \simeq k_\parallel ^{0}  r_g \Omega t$) and we used
\begin{eqnarray}
I(a, u) &=& \int_{u}^{\infty} y^{a-1} {\rm sin} y \,dy \nonumber\\
& = & i/2[e^{-i {\pi\over 2} a} \Gamma(a, iu) - e^{i {\pi\over 2} a} \Gamma(a, -iu)]  
\end{eqnarray}
where $\Gamma (a, z)$ is the incomplete gamma function (see \citet{gr73}, Eq.(3.761.2)). The time evolution of the drift coefficient $d^s_D$ is depicted in Fig.\ref{drift}. The first term in Eq.~(\ref{F1}), corresponding to scales smaller than coherence scale $2\pi/k_\parallel^{min}$ ($ k > k_\parallel^{min}$), dominates over the second term, corresponding to scales larger than $2\pi/k_\parallel^{min}$ ($ k < k_\parallel^{min}$). The diffusive behaviour can be found by using the approximation of $\Gamma(a,z)$ for $|z| = y_\parallel ^{m} = k_\parallel ^{min} v_\parallel t  \ll 1$ (see \citet{gr73}, Eq.(8.354.2)), because $y_\parallel ^{m} \ll 1$; since we assume a weak magnetic fluctuation, it is reasonable to assume that the perpendicular diffusion time-scale is shorter than the parallel scattering time-scale, i.e., $1/k_\parallel ^{min} v_\parallel$, or in other terms the diffusion limit is the dominant term in Eq.~(\ref{F1}) for large $t$ and $t < 1/k_\parallel ^{min} v_\parallel$: $\Gamma(a,z) \sim \Gamma(a) - z^a/a$; therefore $I(2-q, y_\parallel) \sim \sin(q\pi/2)\Gamma(2-q)$ (dashed line in Fig.\ref{drift}). In diffusive regime, the second term in Eq.(\ref{F1}), representing the large scales ($l > 2\pi/k_\parallel^{min}$ or $k < k_\parallel^{min}$), does not significantly contribute to the particle drift, as it is manifest in Fig. \ref{drift}. We find that for slab turbulence, transverse particle drift coefficient from local MFL is given by 
\begin{equation}
\frac{d^s_D (t)}{\kappa_B} \rightarrow {3\over 20} \left( \frac{\delta B}{B_0} \right)^2 \frac{ q-1}{q} \frac{\sin(q\pi/2)\Gamma(2-q)}{(k_\parallel^{min} r_g)^{1-q} (\Omega t)^{2-q}} \, ,
\label{slab_drift_asy}
\end{equation}
thus subdiffusive with behaviour $\kappa^s_D (t) \sim t^{-(2-q)}$ (depicted as the dashed line in Fig.~\ref{drift}).
Transverse subdiffusion has also been found by considering time-scales longer than the parallel scattering time and therefore allowing parallel scattering in \citet{kj00}. However, in that case particles are assumed to propagate back and forth along the MFL and to be tied to the MFL. We notice that the drift-coefficient time evolution in Eq. (\ref{slab_drift_asy}) confirms that charged-particles in a turbulence depending on less than 3 space coordinates remain confined within a gyroradius from the local field line \citep{jkg93, jjb98}.The time-integration of Eq. (\ref{slab_drift_asy}) up to $t =  L_\parallel / v \sim 2\pi/(k^{min}_\parallel r_g \Omega)$, gives in case of weak turbulence ($\delta B \ll B_0$)  the condition $\langle \Delta x ^2 \rangle \ll r_g^2$. We notice that the time-integral of $\kappa^s_D \sim t^{q-2}$, which provides $\langle (\Delta x)^2 \rangle \sim t^{q-1}$, is an increasing function of time for any observed physical value of $q$; however, as shown above, this result does not contradict the theorem of reduced dimensionality. 
In summary, the present result has been obtained under three assumptions: 1) ballistic motion in the $z$ coordinate ($z = v_\parallel t$); 2) average displacement transverse to the local field ${\bf B}$ due to first-order drift; 3) Kolmogorov power spectrum for magnetic fluctuations. 
Equations (\ref{dXXlimit2}, \ref{F1}) represent the average transverse displacement computed in the first-order orbit approximation at any time smaller than the parallel scattering time-scale, so that the approximation of ballistic motion parallel to the mean magnetic field holds. 

\begin{figure}
\includegraphics[width=9cm]{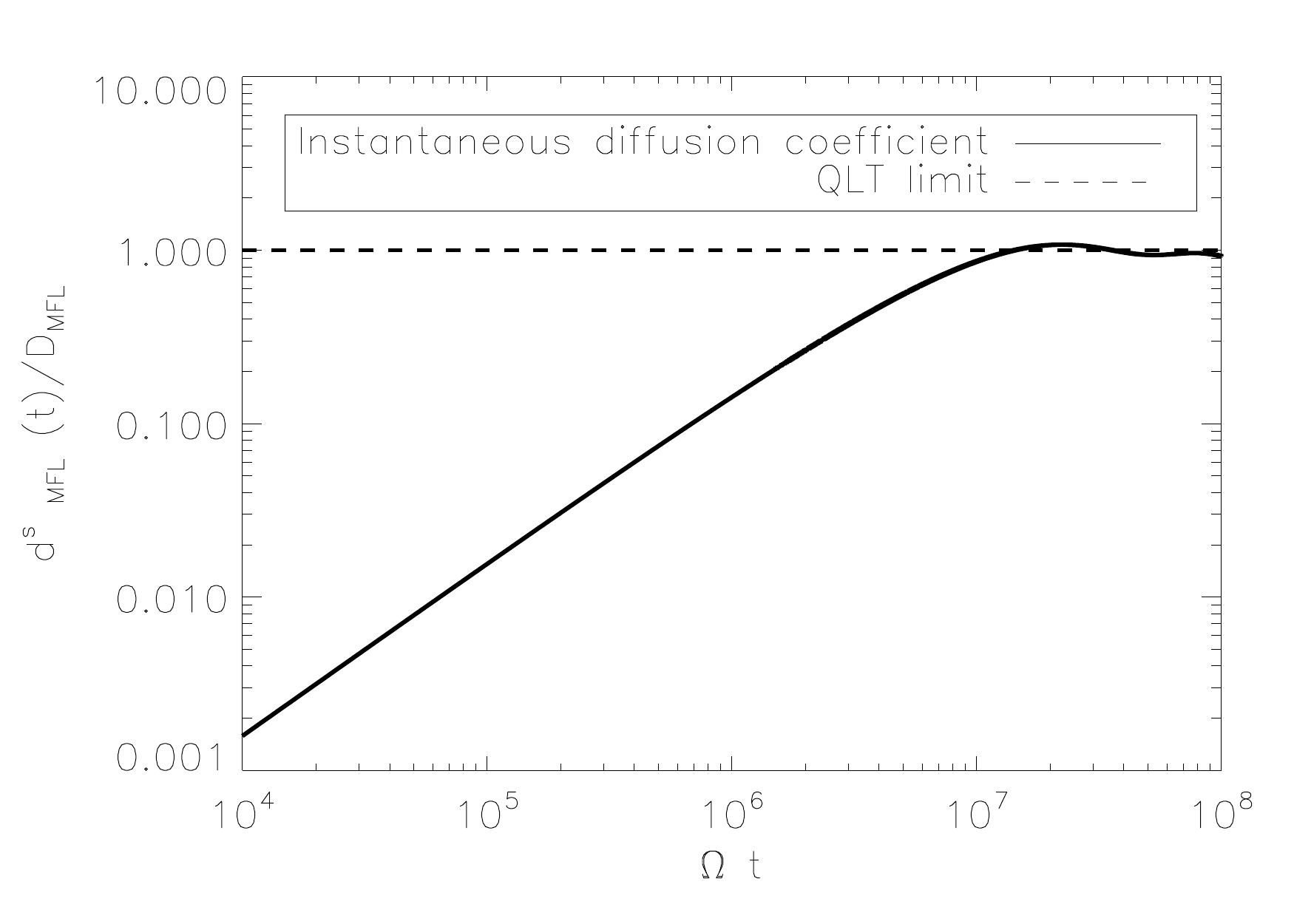}
\caption{Average magnetic field line transverse displacement as computed in Eq.(\ref{dMFL2}) is compared with the QLT coefficient diffusion $D_{MFL}$ as a function of $\Omega t$ for $k_{\parallel} ^{max} r_g= 10^{-3}$, $k_{\parallel} ^{max}/ k_{\parallel} ^{min} =10^4$, $k_{\parallel} ^{min}/k_{\parallel} ^0 =10^2$, $q=5/3$ and $\delta B/ B = 0.1$. The horizontal line represents the quasi-linear limit. The field-line wandering, by computing the magnetic fluctuation along the unperturbed trajectory from times smaller than the time needed to particles to travel the turbulence correlation length, reaches asymptotically the QLT limit. \label{MFL}}
\end{figure}

From Eq.~(\ref{dMFL2}), the MFLRW in units of magnetic coherence length $L_{\parallel} = 2\pi/k_\parallel ^{min}$ is given by
\begin{equation}
d^s _{MFL} (t) k_\parallel ^{min} =   \left( \frac{\delta B}{B_0} \right)^2 \frac{ q-1}{2q} H(y_\parallel,q)
\label{MFL_fin}
\end{equation}
where we defined
\begin{equation}
H(y_\parallel,q) =  (k_\parallel^{min} r_g \Omega t)^q  I(-q, y_\parallel) + {\rm Si}(y_\parallel ^{min}) \, ,
\label{F2}
\end{equation}
where ${\rm Si}(x)$ is the Sine integral function. The first term in Eq.~(\ref{F2}), corresponding to scales smaller than coherence scale $2\pi/k_\parallel^{min}$ ($ k > k_\parallel^{min}$), is dominated by the second term, corresponding to scales larger than $2\pi/k_\parallel^{min}$ ($ k < k_\parallel^{min}$). From Eq.s~(\ref{MFL_fin}, \ref{F2}), the MFLRW diffusion coefficient is given by $\kappa^s_{MFL} = (\delta B/B_0)^2 \pi(q-1)/(4q k^{min}_\parallel)$. In Fig.~\ref{MFL}, the $\kappa^s_{MFL}$ is shown to recover the quasi-linear limit and is dominated by large wavelengths, given by the ${\rm Si}(x)$ term in Eq.~(\ref{F2}): $D_{MFL} = \pi^2 G(k_\parallel = 0)/{B_0}^2 = \kappa^s_{MFL}$, where the quasi-linear limit is expressed, as known, as power spectrum at zero parallel wavenumber.

Equations (\ref{MFL_fin}, \ref{F2}) provide the MFLRW for distances $\Delta z$ along $B_0$ shorter than $L_\parallel$. The guiding center perpendicular scattering in a slab turbulence is described as a series of bumps in the transverse drift which are asymptotically suppressed confining the transverse motion to follow the MFL meandering, as also found in low-rigidity test particle numerical simulations \citep{qmb02}.
Therefore, we confirm that slab transverse transport is due to the meandering of MFLs but we also model the transport across the MFL for first-order magnetic fluctuations not considered in previous treatments \citep{kj00}. For the slab turbulence, the transverse particle diffusion can be disentangled in two energetic contributions: drift coefficient is dominated by length scales smaller than coherence length whereas the MFLRW is dominated by length  scales larger than coherence length.

\subsection{Isotropic}

\begin{figure}
\includegraphics[width=9cm]{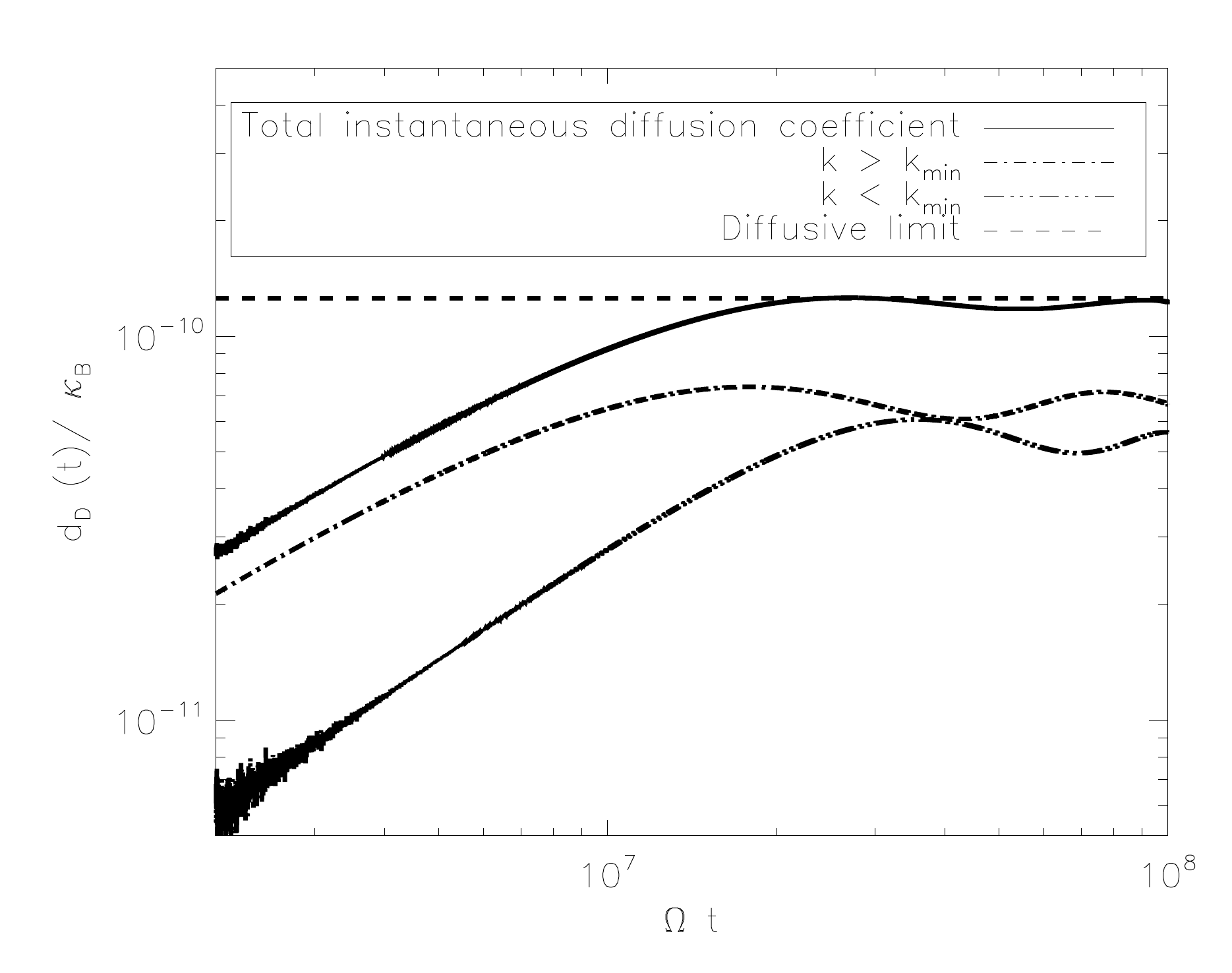}
\caption{Total average transverse drift is compared with the distinct contributions at length-scales smaller and larger than the coherence length in the diffusive regime in units of $\kappa_B$ as a function of $\Omega t$ for $k^{max} r_g= 10^{-3}$, $k_{max}/ k_{min} =10^4$, $k_{min}/k_0 = 10^2$, $q=11/3$ and $\delta B/ B = 0.1$. For comparison the diffusive limit, $\kappa^i_D/\kappa_B$, is shown. In contrast to the slab, length-scales larger than the coherence length give a non-negligible contribution to diffusion. \label{iso}}
\end{figure}

We consider 3D isotropic turbulence, in which the turbulence $\delta B$ depends on all three space coordinates. We adopt the following power spectrum \citep{b70}:
\begin{equation}
P_{rq} ({\bf k}) = \frac{G(k)}{8\pi k^2} \left[ \delta_{lm} - \frac{k_l k_m}{k^2}\right]
\end{equation}
with 
\begin{equation}
 G(k)= \left\{
  \begin{array}{cc}
    G_0 k ^{-q} & \mathrm{if~} k_{min} < k  <  k_{max}  \, \\
    G_0 k_{min} ^{-q} & \mathrm{if~} k_0< k  <  k_{min}\, ,
    \label{spectrumk}
   \end{array}
\right.
\end{equation}
where $k^{max}$ corresponds to the scale where the dissipation rates of the turbulence overcomes the energy cascade rate, the coherence length is given by $L = 2\pi/k^{min}$, and the physical scale of the system by $2\pi/k ^0$ (about the spectrum at large scales, see the discussion in Sect.\ref{slab}).
The constant $G_0$ is fixed by normalization:
\begin{equation}
G_0 = \frac{(\delta B)^2 (q-1)}{q (k^{min})^{1-q}} \, ,
\end{equation}
assuming $k _0 \ll k^{min}  \ll k^{max}$. We use spherical coordinate for the wavenumber  ${\bf k} = k(\sin \theta \cos \psi, \sin \theta \sin \psi, \cos \theta)$ and 3D-turbulence: $\delta {\bf B} ({\bf x}) = (\delta B_x, \delta B_y, \delta B_z)$. We compute first $d_{D_{XX}} ^i (t)$. In reference to Eq.~(\ref{Fmu2}), the non-zero terms are $(r,q,l,p) = (3,3,2,2)$, $(r,q,l,p) = (2,3,2,3)$ and $(r,q,l,p) = (2,2,3,3)$. Combining the non-zero terms and averaging over a pitch-angle isotropic distribution, this gives 
\begin{eqnarray}
\frac{d^i _D (t)}{\kappa_B}  & = &  {3\over 4} \left( \frac{\delta B}{B_0} \right)^2 \frac{ q-1}{q} (k_{min} r_g) \nonumber\\
& \times & \left[\frac{F^i _1(y_m,q)}{y_m^{2-q}} + \frac{F^i _2(y,q)|_{y_0} ^{y_m}}{y_m^{2}}\right]
\label{dXXlimit3}  
\end{eqnarray}
with $y_{m} = k_{min} v_\parallel t \simeq k_{min}  r_g \Omega t$ (and $y_0 \simeq k_{0}  r_g \Omega t$); here, the term integrated over scales smaller than coherence scale $2\pi/k_{min}$ (or $k > k_{min}$) is recast as 
\begin{eqnarray}
F^i _1(y,q) & = & {2\over 15} \frac{-y^{1-q}}{2-q}(\cos y + y {\rm Si}(y)) + {6\over 5} \frac{y^{-q} \sin y}{q} \nonumber\\
& - & {14\over 5} \frac{y^{-2-q} \sin y}{2+q}  + 2 \frac{3 q^2 - 14 q +18 }{15 q(2-q)} \bar I(1-q, y) \nonumber\\
& + & {14\over 5} \frac{1+q}{2+q} \bar  I(-1-q, y)  
\label{F1i}
\end{eqnarray}
and the term integrated over scales larger than coherence scale $2\pi/k_{min}$ (or $k < k_{min}$) is recast as 
\begin{eqnarray}
F^i _2(y,q) & = & \frac{y}{15}(\cos y + y {\rm Si}(y)) - {1\over 5}{\rm Si}(y) \nonumber\\
& + & \sin y  \frac{21 - 5 y^2 }{15 y^2} - {7\over 5} \frac{\cos y}{ y} 
\label{F1i2} 
\end{eqnarray} 
where we approximated $r_g/v_\parallel \simeq \Omega^{-1}$ as in Eq.~(\ref{dXXlimit2}) and we used 
\begin{eqnarray}
\bar I(a, u) &=& \int_{u}^{\infty} y^{a-1} \cos y \,dy \nonumber\\
& = & 1/2[e^{-i {\pi\over 2} a} \Gamma(a, iu) + e^{i {\pi\over 2} a} \Gamma(a, -iu)] \; ,
\end{eqnarray}
from \citet{gr73}, Eq.(3.761.7).
The instantaneous transverse coefficient diffusion in Eq.(\ref{dXXlimit3}) is represented in Fig.\ref{iso}.
The dominant term in the diffusive limit, with the condition $y_{m} = k_{min} v_\parallel t < 1$, is given by
\begin{equation}
\frac{d^i _D (t)}{\kappa_B}  \rightarrow   {\pi\over 40} \left( \frac{\delta B}{B_0} \right)^2 \frac{q-1}{q-2} {(k_{min} r_g)} = \frac{\kappa^i_D}{\kappa_B}
\label{dDiso_asy}
\end{equation} 
and represented in Fig.\ref{iso}. In contrast to the slab, the particle drift from the MFL does not depend only on the power spectrum at length-scales smaller than $L_\parallel$. As for the slab, transverse diffusion is axisymmetric: $d_{D_{XX}} ^i (t) = d_{D_{YY}} ^i (t) = d_D ^i (t)$. Statistically, charged particle motion is not tied to local MFL in 3D isotropic turbulence. Theorem on reduced dimensionality turbulence in \citet{jkg93} and \citet{jjb98} allows charged particle to be magnetized to local MFL within gyroradius scale only in turbulence depending on a reduced number of space coordinates. Any three-dimensional extension could depend on specific geometry but would not be justified in general, as our result shows.

\begin{figure}
\includegraphics[width=9cm]{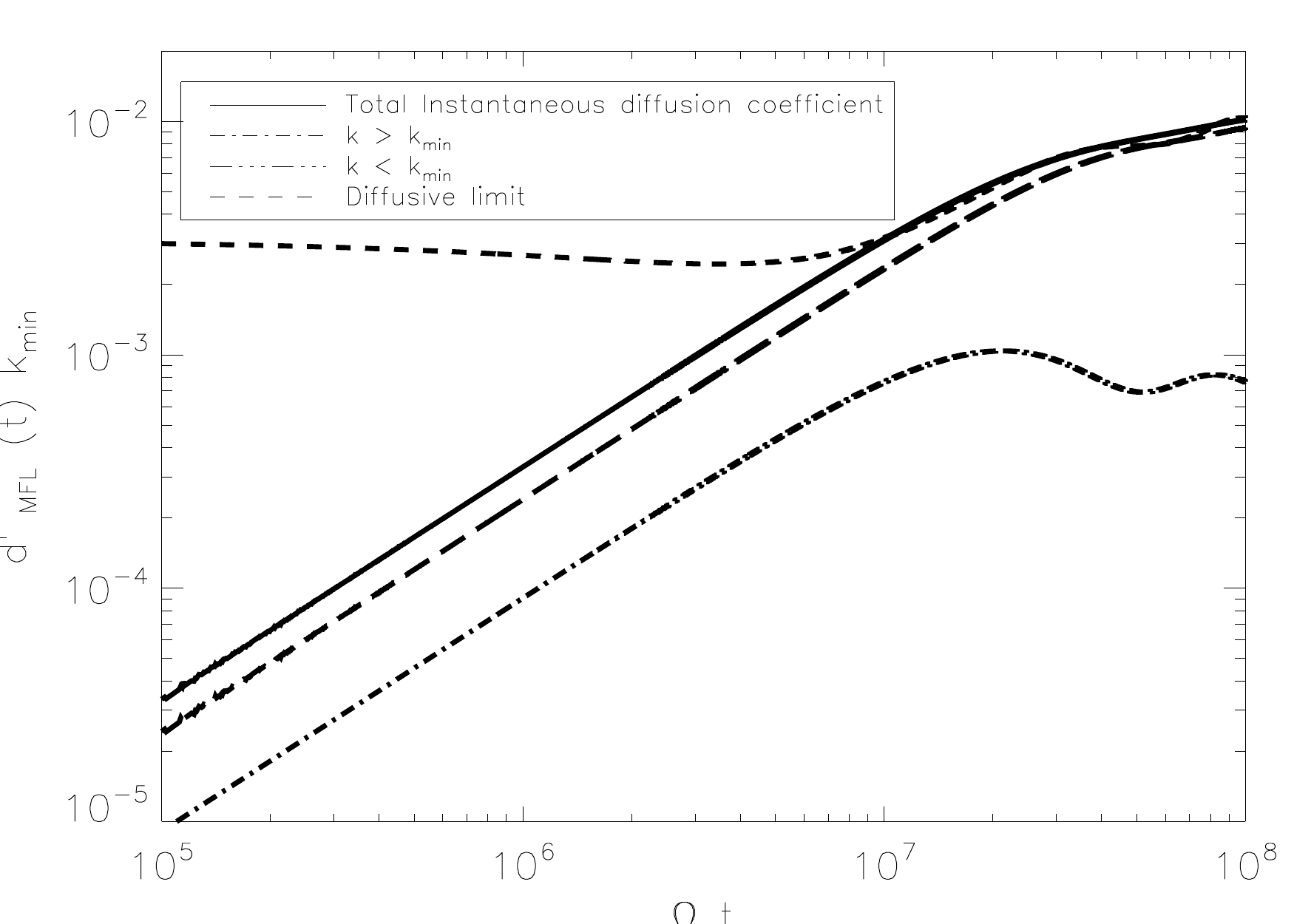}
\caption{Total average magnetic field line transverse displacement for a 3D isotropic turbulence is compared with the distinct contributions at length-scales smaller and larger than the coherence length in the diffusive regime as a function of $\Omega t$ for $k_{max} r_g= 10^{-3}$, $k_ {max}/ k_{min} =10^4$, $k_{min}/k_0 =10^2$, $q=11/3$ and $\delta B/ B = 0.1$. The diffusive limit, $\kappa^i_{MFL} k_{min}$ defined in Eq.~(\ref{MFL_iso_asy}), is overlaid (cf. text).} \label{iso2}
\end{figure}

Comparison with previous numerical simulations (see, e.g., \citet{gj99}) requires the evaluation of the MFLRW, for isotropic turbulence. Using Eq.~(\ref{dMFL2}), we find for the average square displacement of MFL (same contribution along $x$- and $y$-axes)
\begin{equation}
d^i _{MFL} (t) k_{min}  =   {1\over 4} \left( \frac{\delta B}{B_0} \right)^2 \frac{ q-1}{q} \left[ H^i _1(y_m,q)+ H^i _2(y_m,y_0,q) \right] \, ;
\label{MFLtot}
\end{equation}
here the term integrated over scales smaller than coherence scale $2\pi/k_{min}$ (or $k > k_{min}$) is recast as 
\begin{eqnarray}
H^i _1(y,q) & = & \frac{1}{q y}(\cos y + y {\rm Si}(y))  + \frac{y^{-2}}{2+q}\sin y \nonumber\\
&- & 2\frac{q^2 + 2q +1}{q(2+q)} y^q I(-1-q, y)
\label{H1i}
\end{eqnarray}
and the term integrated over scales larger than coherence scale $2\pi/k_{min}$ (or $k < k_{min}$) is recast as 
\begin{eqnarray}
H^i _2(y_m,y_0,q) & = & \int_{y_0} ^{y_m} dy \frac{{\rm Si}(y)}{y} + \left( \frac{{\rm Si}(y)}{2} - \frac{\sin y}{2 y^2} + \frac{\cos y}{2 y} \right) \Bigl\lvert_{y_0}^{y_m}\nonumber  \, . \\
\label{H2i}
\end{eqnarray}
Figure ~\ref{iso2} shows the diffusive behaviour of MFL for a magnetic turbulence with isotropic wave-number spectrum. As for the slab case, scales larger than coherence length ($k<k_{min}$) dominate over the turbulent contribution (see Fig.~\ref{iso2}). The leading terms in the diffusive limit of $H^i _2(y,q)$ are found in the integral and in the first term in parenthesis of Eq.~(\ref{H2i}). By integrating by parts and using \citet{gr73}, Eq.(4.421.1), it can be found 
\begin{eqnarray}
d^i _{MFL} (t) k_{min} & \rightarrow &  {1\over 4} \left( \frac{\delta B}{B_0} \right)^2 \frac{ q-1}{q} \left[{\rm Si}(y){\rm log}(y)|^{y_m}_{y_0} +  {\pi \over 2} \left({\cal C} + {1\over2}\right) \right] \nonumber\\
& = & \kappa^i_{MFL} k_{min}  \, ,
\label{MFL_iso_asy}
\end{eqnarray}
where ${\cal C} = 0.577215$ is the Euler constant and $y_{m} = k_{min} v_\parallel t \simeq k_{min}  r_g \Omega t$ ($y_0 \simeq k_{0}  r_g \Omega t$). 
We find that the MFL of a 3D isotropic weakly turbulent magnetic field are superdiffusive 
according to Eq.~(\ref{MFL_iso_asy}). This result does not imply the superdiffusion of the particles which propagate diffusively in a 3D isotropic turbulence \citep{gj99}. Before being transported superdiffusively along a field line, a particle will undergo parallel scattering, not taken into account in this paper, and eventually decorrelate to another field line.
These results might be qualitatively extended to MHD-turbulence with a small k-anisotropy so that   results in diffusion regime found here still apply. In this case, particle drift retains its diffusive character shown in Fig.~\ref{iso}, even if only scales smaller than the coherence length ($k > k_{min}$, in dotted-dashed) are taken into account. Therefore, at small length-scales, the perturbative approach based on $\delta B \ll B_0$ could be applied to solar wind turbulence. At length-scales much larger than the coherence scale ($k \ll k_{min}$), where the approximation $\delta B \ll B_0$ breaks down in the solar wind, these conclusions cannot be extended. 

\section{Discussion and conclusion}

We have described analytically the time evolution of individual charged particles drift and MFLRW across a static magnetic field to first-order in the  magnetic fluctuations. We consider the case where the motion perpendicular to the average magnetic field is dominated by guiding center motion which includes the meandering of the MFL and the drift from the first-order orbit theory, in the approximation that the particle gyroradius is much smaller than the length scale of magnetic field variations. In contrast to previous models for the perpendicular transport, we do not assume diffusive scattering {\it at all times}; this allows us to treat consistently the slab turbulence perpendicular diffusion. Drift and MFL transverse transport are explicitly computed for both the slab and 3D isotropic cases. In the slab case, the time-evolution of the drift displacement shows how the particle diffusion from the MFL is suppressed. 
The instantaneous slab drift coefficient diffusion transverse to the local field depends on the turbulence power-law spectral index; for a Kolmogorov turbulence is found to decrease as $t^{-1/3}$, slower than compound diffusion displacement ($t^{-1/2}$), which is however computed transversally to the average field and not to the local magnetic field as in this paper. The recovery of the MFL coefficient diffusion of QLT shows that this result does not depend on assuming MFL parallel diffusion. Secondly, we provide analytical time-dependence of drift and MFL coefficients of diffusion for a 3D isotropic turbulence. We found that for a 3D isotropic turbulence the particle drift from the local field line is diffusive, whereas the field line itself is superdiffusive. Previous numerical simulations are not contradicted by our result which is obtained neglecting scattering parallel to the mean field.
For the slab, we find that MFLRW is dominated by length-scales larger than the coherence length whereas particle drift from the local field line is dominated by  length-scales smaller than the coherence length. This disentaglement does not hold for 3D isotropic turbulence: MFLRW is still dominated by large length-scales whereas as far as drift is concerned turbulent energy contribute at all scales, both below and above the coherence length. The study carried out here provides a framework for particle transport in the solar wind and supernova remnant blast wave turbulence and questions the common assumption that cosmic-rays trajectory follow the magnetic field line.

\acknowledgments

It is a pleasure to acknowledge the fruitful discussions with J. Giacalone, J. K\'ota and D. Ruffolo. 
We thank the anonymous referee for useful suggestions and comments. 
The work of FF was supported by NSF grant ATM0447354 and by NASA grants NNX07AH19G 
and NNX10AF24G; the work of JRJ was partially supported by NASA grant NNX08AH55G.

\clearpage

\end{document}